\documentclass[12pt,a4paper,dvips]{article}

\usepackage{graphicx}
\usepackage{amsbsy}
\usepackage{amssymb}

\begin{document}

\newcommand{\be}{\begin{equation}}
\newcommand{\ee}{\end{equation}}
\newcommand{\bea}{\begin{eqnarray}}
\newcommand{\eea}{\end{eqnarray}}
\newcommand{\tr}{\,\hbox{tr }}
\newcommand{\Tr}{\,\hbox{Tr }}
\newcommand{\Det}{\,\hbox{Det }}
\newcommand{\fslash}{\hskip-.2cm /}
\begin{center}
{\bf \Large The Measure from Schwinger-Dyson Equations}
\end{center}
\vspace*{0.2cm}
\begin{center}
A. Bogojevi\'c and D. S. Popovi\'c
\end{center}
\begin{center}
{\it Institute of Physics\\
P.O.B. 57, Belgrade 11001, Yugoslavia}
\end{center}
\vspace*{0.1cm}
\begin{abstract}
{\small We review a new prescription for calculating the Lagrangian 
path integral measure directly from the Hamiltonian Schwinger-Dyson 
equations. The method agrees with the usual way of deriving the measure 
in which one has to perform the path integration over momenta.}
\end{abstract}
\vspace*{0.1cm}

Linearity (of amplitudes) lies at the heart of any quantum theory. In
the formalism of quantum field theory this linearity is encoded in the 
Schwinger-Dyson equations. In terms of the generating functional $Z[J]$ 
we have
\be
\left({\delta S\over\delta q}
\Bigm|_{q={\hbar\over i}\,{\delta\over\delta J}}+J\right)
Z[J]=0\ ,\label{sd}
\ee
where $S[q]$ generates the Feynman rules: $S''$ is the inverse of the
Feynman propagator, $S^{(n)}$ are the $n$-point vertices. Equation
(\ref{sd}) is a linear (functional) differential equation for $Z[J]$. 
A formal functional Fourier transform of this is just the Feynman path 
integral
\be
Z[J]=\int [dq]\,e^{\frac{i}{\hbar}\left(S[q]+\int dt\,Jq\right)}
\ .\label{path-integral}
\ee
The $\hbar\to 0$ limit of $Z[J]$ is dominated by configurations near to the
solutions of
${\delta S\over\delta q}+J=0$. On the other hand,
$\hbar\to 0$ corresponds to classical physics given by
${\delta I\over\delta q}+J=0$, where $I$ is the action. From this we see
that
\be
S[q]=I[q]+{\hbar\over i}\,M[q]\ .
\ee
$M[q]$ is the measure term. This is as far as the usual functional formalism
takes us --- namely, there is no way to determine the measure
term\footnote{It is common to write the integrand of (\ref{path-integral})
in terms of the action. The path integral measure is then 
$d\mu=[dq]\,\exp M[q]$. Lack of knowledge about $M$ translates into lack of
knowledge about $d\mu$.}.
The only way to do this has been to make connection with the operator
formalism. From it we find an expression for the generating functional in
terms of a Hamiltonian path integral
\be
Z[J]=\int [dp\,dq]\exp{i\over\hbar}\,\int dt\,
\big(p\dot q-H(q,p)+J q\big)\ .
\ee
Here the measure is trivial. The Lagrangian expressions, including
the corresponding measure, are obtained by doing the momentum path integral
(see for example \cite{iz}).

In a previous paper \cite{bp} we have developed an alternate way for
calculating the measure \emph{inside} the functional formalism. To do this
we start from the Hamiltonian form of the Schwinger-Dyson equations. 
These may be written as
\bea
\left(\dot P+{\partial H(Q,P)\over\partial Q}-J \right)Z[J ,K]
&=& 0\\
\left(\dot Q-{\partial H(Q,P)\over\partial P}+K\right)Z[J ,K]
&=& 0\ ,
\eea
where $K$ is an additional source term that couples to momenta\footnote{
The corresponding Hamiltonian path integral is 
$$
Z[J ,K]=\int [dp\,dq]\,\exp{i\over\hbar}\int dt\,
\left(p\dot q-H(q,p)+J q+Kp\right)\ ,
$$
where the $[dp\,dq]$ measure is trivial.}, 
$P={\hbar\over i}{\delta\over\delta K}$, and
$Q={\hbar\over i}{\delta\over\delta J }$. With this nomenclature the 
above Schwinger-Dyson equations look just like the classical Hamiltonian 
equations of motion. The only difference is that we have the following 
non-zero commutators
\be
[P,K]=[Q,J]={\hbar\over i}\ .
\ee
Note that in this formalism $P$ and $Q$ commute. By using the above
equations we can derive the Lagrangian path integral measure. 
As an example let us look at a model whose Hamiltonian is simply
\be
H(q,p)={1\over 2}\,p^2+V(q)\ .\label{simple-hamiltonian}
\ee
In this case the Schwinger-Dyson equations read
\bea
\left(\dot P+V'(Q)-J \right)Z[J ,K] &=& 0\\
\left(\dot Q-P+K\right)Z[J ,K] &=& 0\ .
\eea
Differentiating the second of these equations with respect to time,
and then adding it to the first, we get an equation for $Q$ alone
\be
\left(\ddot Q-V'(Q)-J\right)Z[J]=0\ ,
\ee
where we have now turned off the source for momenta. The action for this
model is $I[q]=\int dt\,\big({1\over 2}\,\dot q^2-V(q)\big)$. In terms of it
the Schwinger-Dyson equations are simply
\be
\left({\delta I\over\delta Q}+J\right)Z[J]=0\ .
\ee
Fourier transforming this we obtain the usual Lagrangian path integral
\be
Z[J]=\int [dq]\,\exp{i\over\hbar}
\int dt\,\left({1\over 2}\,\dot q^2-V(q)+Jq\right)\ .
\ee
We have just derived the well known result that the path integral measure 
is trivial for models whose Hamiltonian is of the simple form given
in (\ref{simple-hamiltonian}). 

Now let us look at a bit more complicated example. We consider a
model with Hamiltonian given by
\be
H(q,p)={1\over 2}\,g^{-1}(q)p^2+V(q)\ .
\ee
The Hamiltonian Schwinger-Dyson equations are now
\bea
\left(\dot P-{1\over 2}g^{-2}(Q)g'(Q)P^2+V'(Q)-J \right)Z[J ,K]
&=& 0\\
\left(\dot Q-g^{-1}(Q)P+K\right)Z[J ,K]
&=& 0\ .
\eea
We may write the second equation as $PZ=g\,(\dot Q+K)Z$ and use this to
get rid of the $P$ terms in the first equation. Therefore
\bea
\lefteqn{
P^2Z=P(g\dot Q+gK)Z=}\nonumber\\
& &{}=(g\dot Q+gK)PZ+[P,K]gZ=
\left((g\dot Q+gK)^2+{\hbar\over i}\,g\right)Z\ ,
\eea
as well as
\be
\dot PZ=\big(g'\dot Q(\dot Q+K)+g(\ddot Q+\dot K)\big)Z\ .
\ee
Setting $K=0$ we find
\be
\big(g\ddot Q+{1\over 2} g'\dot Q^2 + V'-{1\over 2}\,{\hbar\over i}\,(\ln g)'
-J \big)Z[J ]=0\ .
\ee
This equation can again be written as
$\left({\delta S\over\delta Q}+J \right)Z[J]=0$,
where we have $S=I+{\hbar\over i}M$. The first term is just the action
$I[q]=\int dt\,\big(\,{1\over 2}g(q)\dot q^2-V(q)\big)$. The
measure term equals $M=\int dt\,\ln\sqrt{g}$. Fourier transforming the last
equation we find
\be
Z[J ]=\int\prod_t\left(dq(t)\sqrt{g(q)}\right)
\exp\,{i\over \hbar}\big(I+\int dt\,J q\big)\ .
\ee
This agrees with the standard derivation of the Lagrangian
path integral in which one performs the Gaussian momentum integration in the 
Hamiltonian path integral. 

The generalization of the previous example to more variables gives us
the $\sigma$-model
\be
L={1\over 2}\,g_{\alpha\beta}(q)\dot q^\alpha\dot q^\beta\ .
\ee
The Hamiltonian is given in terms of the inverse metric
$g^{\alpha\beta}$, and equals
$H={1\over 2}\,g^{\alpha\beta}p_\alpha p_\beta$. The Schwinger-Dyson
equations become
\bea
\left(\dot P_\alpha+{1\over 2}g^{\gamma\delta}_{\ ,\alpha}P_\gamma P_\delta
-J _\alpha\right)Z[J ,K] &=& 0\\
\left(\dot Q^\alpha-g^{\alpha\beta}P_\beta+K^\alpha\right)Z[J ,K] 
&=& 0\ .
\eea
Just as in the previous example it is a simple exercise to get rid of
the $P$ terms and derive the Lagrangian Schwinger-Dyson equation. It may be
compactly written as
\be
\left({\delta I\over\delta Q^\alpha}-i\hbar\,
{1\over\sqrt{g}}\,\partial_\alpha\sqrt{g}+J_\alpha\right)Z[J ]=0
\ ,
\ee
where $g=\hbox{det }g_{\alpha\beta}$. The corresponding path integral
has the familiar form
\be
Z[J ]=\int\prod_t\left(dq(t)\sqrt{g(q)}\right)
\exp\left(\,{i\over \hbar}\big(I+\int dt\,J _\alpha q^\alpha\big)\right)
\ .
\ee
From these examples it is obvious that the generalization from
1-dimensional field theory, i.e. quantum mechanics, to $d$-dimensional field
theory is trivial. The $d$-dimensional expressions just contain more
dummy labels. What is not trivial, when one tackles full-fledged field theory, 
is how to deal with gauge symmetries. Therefore, it would be very 
interesting to extend this work to the treatment of gauge theories,
and re-derive the measures obtained by Faddev-Popov and Batalin-Vilkovisky.
Another interesting avenue of research is to try to use the above 
method to derive an explicit differential equation satisfied by the
measure term. Doing this would enable us to complete what Dirac and 
Feynman started: To define a complete quantum theory in terms of the
Lagrangian.

\end{document}